# Brillouin-Kerr soliton frequency combs in an optical microresonator

Yan Bai[1]†, Menghua Zhang[1]†, Qi Shi[1]†, Shulin Ding[1]†, Zhenda Xie[1], Xiaoshun Jiang[1]*, and Min Xiao[1,2]

[1]National Laboratory of Solid State Microstructures, College of Engineering and Applied Sciences, and School of Physics, Nanjing University, Nanjing 210093, China.
[2]Department of Physics, University of Arkansas, Fayetteville, Arkansas 72701, USA.

**Abstract:** By generating a Brillouin laser in an optical microresonator, we realize a soliton Kerr microcomb through exciting the Kerr frequency comb using the generated Brillouin laser in the same cavity. The intracavity Brillouin laser pumping scheme enables us to access the soliton states with a blue-detuned input pump. Due to the ultra-narrow linewidth and the low-noise properties of the generated Brillouin laser, the observed soliton microcomb exhibits narrow-linewidth comb lines and stable repetition rate even by using a diode laser with relatively broad linewidth. Also, we demonstrate a low-noise microwave signal with phase noise levels of -24 dBc/Hz at 10 Hz, -111 dBc/Hz at 10 kHz, and -147 dBc/Hz at 1 MHz offsets for a 11.14 GHz carrier with only a free-running input pump. The easy operation of the Brillouin-Kerr soliton microcomb with excellent performance makes our scheme promising for practical applications.

Owing to the advantages of the soliton frequency comb [1,2] in microresonators such as broad spectrum span, high repetition rate as well as on-chip integration, the soliton microcombs have been widely used in coherent optical communications [3], low-noise microwave signal generation [4-6], dual-comb spectroscopy [7,8], ultrafast optical ranging [9,10], astronomical spectrometer calibration [11,12], optical coherence tomography [13], optical frequency synthesizer [14] and optical atomic clock [15]. Although the soliton microcombs have many promises and have been achieved in several material platforms [1,2,16], it is still challenging to stably approach the soliton states of the Kerr microcomb for practical applications, due to the large thermo-optic nonlinearity in high-Q microcavity and red-pump-detuning requirement [17] for soliton generation. So far, a number of schemes including power kicking [18,19], fast thermal tuning [20], pulsed laser driving [21], single sideband modulation [22] and using an auxiliary laser [23-25] have been used to fast tune the laser frequency or compensate the thermo-optic effect in high-Q optical microresonators. Unfortunately, most of those methods require additional optical and electrical components, which make the soliton generation complicated and difficult for full on-chip integration.

In addition to the parametric nonlinearity that is used for Kerr soliton microcomb generation [1,2], stimulated Brillouin scattering (SBS) [26] is another interesting nonlinear optical effect in high-Q optical microresonators. In previous work, due to the high gain and narrowband of the SBS process, the SBS effect in microcavity has been used for producing on-chip ultra-narrow linewidth laser [27,28], which further leads to the applications in Brillouin gyroscope [29] and low noise microwave signal generation [30]. Despite the fact that the simultaneous presences of Brillouin laser and parametric oscillation were demonstrated in fiber Fabry-Pérot (FP) cavity [31], microbottle [32] as well as microbubble cavities [33], yet no soliton microcomb with Brillouin lasing has been observed so far.

Here, we report a new approach for efficient generation of the soliton microcomb in a high-Q microresonator by producing a narrow linewidth Brillouin laser first and then using this intracavity Brillouin laser to generate the Kerr frequency comb in the same cavity [Fig. 1(a)]. In contrast to the Stokes soliton [34], here the Brillouin process experiences gain with a much narrower bandwidth and the pump light does not need to create a soliton frequency comb. By properly selecting the mode spacing between the pump and Brillouin modes in combination with the Kerr self-phase modulation, we can achieve a red-detuned Brillouin laser with a blue-detuned input pump, which results in long soliton steps and enables stable access to the single soliton state by simply tuning the laser piezo. In addition, this approach is capable of self-stabilizing the system because of the thermo-optic nonlinearity [35]. This new type of soliton Kerr microcomb features narrow linewidth comb lines and stable repetition rate. Besides, in this work we have further realized a low-noise microwave signal based on the generated Brillouin-Kerr soliton microcomb.

In the experiment, a 6 mm-diameter silica microdisk resonator [Fig. 2(a)] with a thickness of 8 μm was fabricated using the previously established method [36]. The wedge angle of the fabricated silica microdisk is around 10°, which is due to the over-etching of the buffered HF etching to reduce the surface roughness of the side wall. The measured intrinsic Q-factors for the Brillouin mode and the pump mode are measured to be $6.33 \times 10^7$ [Fig. 2(b)] and $4.01 \times 10^7$, respectively. As shown in Fig. 2(c), the measured free spectral range of the Brillouin mode is 12.02 GHz, and the mode spacing between the pump mode and the Brillouin mode used for the Brillouin-Kerr soliton mcirocomb generation is 11.25 GHz.

In order to generate the Brillouin-Kerr frequency comb, the Brillouin mode and its corresponding mode family should exhibit anomalous group velocity dispersion while the pump mode and its corresponding mode family do not need to satisfy this requirement. In our experiment, only the mode family of the Brillouin mode possesses anomalous group velocity dispersion, which makes the Kerr frequency comb only occur in this mode family. Also, for the formation of a bright soliton, the generated Brillouin laser should be red detuned relative to the cavity mode [17,37], which can be ensured by the Kerr self-phase modulation effect of the Brillouin laser with a certain mode space between the pump and the Brillouin modes (see Supplementary Material [38], Section II). In the experiment, we first set the input pump laser to a relatively high power (much higher than the threshold) and scan the laser frequency with a speed of 349 MHz/ms. During the pump frequency tuning, the Brillouin lasing will occur first with a far blue-detuned input pump [Figs. 2(d) and 2(e)]. Figures 2(f) and 2(g) show, respectively, the measured optical spectrum of the Brillouin laser and RF spectrum of the beat note between the backscattered pump laser and the generated Brillouin laser with a frequency of 10.78 GHz before the formation of the Brillouin-Kerr comb. As shown in Figs. 2(d) and 2(e) by gradually decreasing the frequency of the input pump, we can observe distinct soliton steps of the Brillouin-Kerr comb when the input pump laser is blue detuned (relative to the cavity resonance), which indicates that the generated Brillouin laser is red detuned with respect to the Brillouin cavity mode [17]. Since the frequency of the generated Brillouin laser is insensitive to the fluctuation of input pump laser frequency [39] (also see Supplementary Material [38], Section II), the experimentally obtained soliton steps are significantly broadened to dozens of MHz, which is much wider than that achieved in the previous work [18] with a silica microdisk resonator. The observation of the soliton steps with a blue-detuned input pump (although the Brillouin laser is red detuned) implies that the generated Kerr solitons can be thermally self-stable [35] even with the red-detuned Brillouin laser excitation (see Supplementary Material [38], Sections IV and V). Moreover, the interaction between the input pump and the generated Brillouin laser will also alter the cavity lineshape of the pump mode [39,40], which extends the thermal self-stability of the pump into the red-detuning region (see Supplementary Material [38], Section IV).

To see how the generated Brillouin laser is red-detuned and the soliton steps are broadened, we consider the phase matching condition of the SBS process in a Kerr nonlinear cavity and obtain the frequency detuning of the Brillouin laser (see Supplementary Material [38] for more details, Section II):

$$\Delta\omega_{b,0} = \Delta\omega'_{b,0} + \Delta\omega''_{b,0}, \tag{1}$$

where $\Delta\omega'_{b,0} = -\frac{\gamma_m}{\gamma_m+\gamma_-}g_2|a_{-,0}|^2$ and $\Delta\omega''_{b,0} = \frac{\gamma_-}{\gamma_m+\gamma_-}(\Delta\omega_0 + \Delta\omega_p)$. Here, $\Delta\omega_{b,0} = \omega_{b,0} - \omega_{-,0}$ and $\Delta\omega_p = \omega_p - \omega_+$ represents pump laser detuning, with $\Delta\omega_0 = \omega_+ - \omega_{-,0} - \Omega_0$ being the resonant frequency mismatch among the three modes: $\omega_+$ (pump), $\omega_{-,0}$ (Brillouin) and $\Omega_0$ (acoustic). $\omega_p$ and $\omega_{b,0}$ are the pump and Brillouin laser frequencies, respectively. $\gamma_m$ and $\gamma_-$ represent the decay rates of the acoustic and Brillouin modes, respectively, $g_2$ is the Kerr nonlinear coupling coefficient, and $a_{-,0}$ is the amplitude of the Brillouin laser. As shown in Eq. (1), the frequency shift of the Brillouin laser originates from two contributions, $\Delta\omega'_{b,0}$ and $\Delta\omega''_{b,0}$ [Fig. 1b], which are, respectively, determined by the Kerr self-phase modulation and the SBS processes. According to a previous study [17, 37], the generation of the bright Kerr solitons require a red-detuned pump. In the case of our Brillouin-Kerr solitons, this requirement implies $|\Delta\omega'_{b,0}| > \Delta\omega''_{b,0}$. On one hand, in the limit of $\gamma_m \gg \gamma_-$, we find $\Delta\omega'_{b,0} \approx -g_2|a_{-,0}|^2$ and $\Delta\omega''_{b,0} \approx \frac{\gamma_-}{\gamma_m}(\Delta\omega_0 + \Delta\omega_p)$. As a result, for $\Delta\omega'_{b,0}$, we expect the red shift of the resonant Brillouin mode due to the self-phase modulation would give rise to an equivalent red shift on the Brillouin laser [Fig. 1b]. In this way, it makes a red-detuned Brillouin laser feasible easily in the cavity. On the other hand, $\Delta\omega''_{b,0}$ is determined by the frequency mismatch among the three modes plus the pump frequency detuning weighted by a small factor $\frac{\gamma_-}{\gamma_m} \ll 1$. This suggests that the large pump detuning modulation merely gives a small frequency shift for the Brillouin laser, because of the relatively small variation of $|a_{-,0}|^2$. With these, we conclude that this property is crucial for broadening the soliton steps when scanning the frequency of the input pump.

Owing to the wide soliton steps of the Brillouin-Kerr soliton microcomb, we can gradually access to the soliton states by manually adjusting the piezo of the input laser. In the measurements, we first fix the optical power and gradually decrease the input pump laser frequency. By tuning the frequency of the input pump laser into the soliton steps, multi-soliton states can then be generated subsequently. After that, we use the backward-tuning method [41] to reduce the number of solitons until the single soliton is observed. As shown in Fig. 3(a), the blue curves represent the measured optical spectra of different soliton states with a repetition rate of 11.14 GHz. The orange curves show the fitting curves for the corresponding soliton states using the method described in Ref. [42]. The orange curve in III of Fig. 3(a) gives the single-soliton envelope of $sech^2$ fitting for a soliton pulse width of approximately 461 fs. The deviations between the fitting curves and the measured optical spectra stem from the Raman self-shift [43,44] and the avoided mode crossing [45]. Also, the measured linewidth of the generated microwave signal [Fig. 3(b)] is as narrow as 10 Hz which is much narrower than the previously reported results [17,18]. To explain the generation of the Brillouin-Kerr soliton microcomb, we have developed a theoretical model (see Supplementary Material [38], Section I and V) by fully taking into account the nonlinear FWM and SBS processes, as well as the interaction between them, which matches well with our experimental results.

Due to the narrow linewidth of the Brillouin laser generated in high-Q microresonator [27,28] that can correspondingly produce Kerr frequency comb with narrow-linewidth comb lines [46] and the thermal stability of the Brillouin-Kerr soliton, the demonstrated microcomb here allows us to produce low noise microwave signal with a free-running pump laser. In the previous works, both the Brillouin laser [30] and Kerr soliton microcombs [4-6,47] have been demonstrated for the generation of low-noise microwave signal. Those works typically employed active lock approaches, such as injection locking [5,6,47], Pound-Drever-Hall (PDH) locking [5,6,47], and phase-lock loop [30], which require relatively complex optical and electrical components in order to stabilize the Kerr soliton. In contrast, here by combining the advantages of both the Brillouin laser and the soliton microcomb generated in the same high-Q cavity, we realize a low-noise microwave signal generation using only a free-running pump laser. As one noticeable advantage, this setup makes the chip-integration of the system easy for practical microwave photonics applications. In the experiment, to measure the phase noise of the soliton's repetition rate, we filtered out the reflected pump laser with an FBG fiber filter [Fig. 1(c)] and photo-mixed the frequency comb with a fast photodiode. Figure 4(a) shows the measured phase noise of the repetition frequency of the Brillouin-Kerr soliton in the region of a multi-soliton state. The obtained phase noise levels are, respectively, -24 dBc/Hz at 10 Hz, -49 dBc/Hz at 100 Hz, -68 dBc/Hz at 1 kHz, -111 dBc/Hz at 10 kHz, -131 dBc/Hz at 100 kHz, and -147 dBc/Hz at 1 MHz. Comparing to previously demonstrated soliton microcombs with a silica microdisk resonator [18] (where a narrow-linewidth fiber laser was used as the pump and its detuning was locked to stabilize the soliton), in our current work, the obtained phase noise is much lower even with a free-running pump laser. As one can see, the phase noise is well kept around 10 dB lower in nearly all the offset frequencies despite the fact that the employed pump laser (Toptica CTL 1550) has a relatively broader linewidth. The frequency stability of the repetition rate is characterized by the Allan deviation with a gate time 0.1 s, as shown in Fig. 4(b). The stability is manifested by the Allan deviations at the level of $5.0\times10^{-9}$ at 1 s and $4.5\times10^{-9}$ for 100 s average time, which are much lower than the previous results in soliton microcomb in a silica microresonator with a free-running pump [48] or Brillouin microwave oscillator with open-loop operation [30].

In conclusion, we have demonstrated a Brillouin-Kerr soliton microcomb with narrow-linewidth comb lines and stable repetition rate. In contrast to the previous work with a red-detuned pump, our approach with blue-detuned pump makes the system be thermally self-stabilized. We have also obtained the ultra-low noise microwave signal based on this soliton microcomb with only a free-running pump laser. Our scheme should also be applicable to other material platforms such as crystal and silicon nitride resonators [28,49], where both Kerr soliton and Brillouin nonlinearity have already been separately demonstrated. In particular, our approach can be adapted in the Mid-IR wavelength range where the narrow linewidth pump laser is limited. This new kind of Kerr soliton microcomb can find applications in optical atomic clock, dual-comb spectroscopy, coherent optical communication, astrocomb, as well as dual-comb ranging. Moreover, turn-key operation [50] for such thermally self-stabilized Kerr soliton microcomb from the current scheme will be beneficial for future applications.

**Acknowledgments:** We thank Prof. Kerry Vahala for helpful comments and suggestions on this work. This research was supported by the National Key R&D Program of China (2017YFA0303703, 2016YFA0302500), the National Natural Science Foundation of China (NSFC) (61922040, 11621091), and the Fundamental Research Funds for the Central Universities (021314380170).

*Email: jxs@nju.edu.cn.

†These authors contributed equally：Yan Bai, Menghua Zhang, Qi Shi, Shulin Ding.

**Figures:**

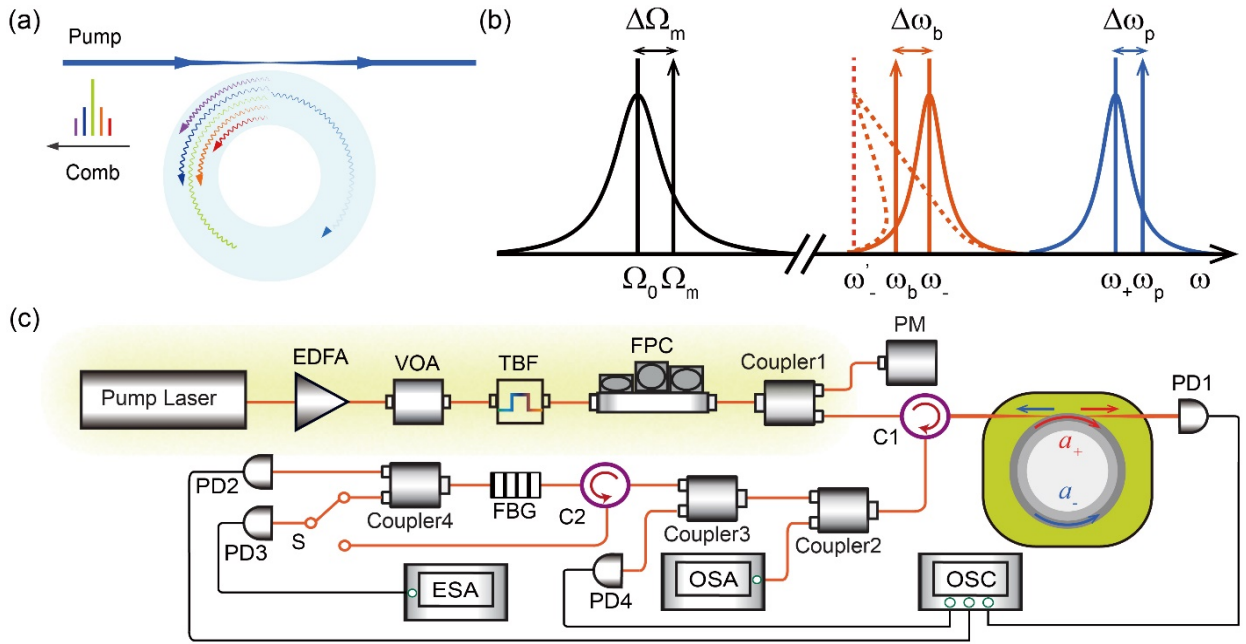

FIG. 1. Scheme of Brillouin-Kerr soliton frequency comb in a silica microresonator. (a) Schematic diagram of Brillouin-Kerr soliton frequency comb. (b) Illustration of red-detuned Brillouin laser with Kerr self-phase modulation. Dash orange curve, cavity response of Brillouin mode with Kerr self-phase modulation. See more details in Fig. S1 of the Supplementary Material [38]. (c) Experimental setup for generation and characterization of the Brillouin-Kerr soliton frequency comb. EDFA, erbium-doped fiber amplifier; VOA, variable optical attenuator; TBF, tunable bandpass filter; FPC, fiber polarization controller; PM, power meter; C, optical fiber circulator; PD, photodiode; FBG, fiber Bragg grating; S, optical switch; OSC, oscilloscope; OSA, optical spectrum analyzer; ESA, electric spectrum analyzer.

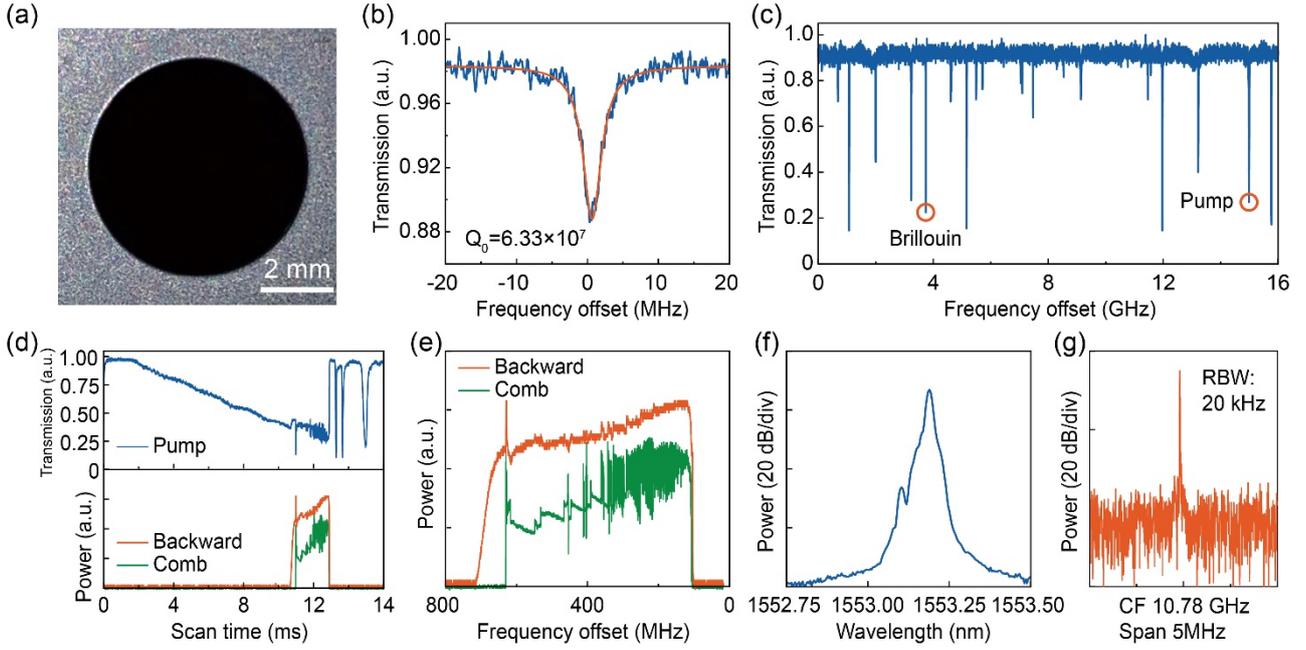

FIG. 2. Characterizations of the microdisk resonator, Brillouin laser and Brillouin-Kerr soliton frequency comb. (a) Photograph of the 6-mm-diameter disk resonator. (b) The transmission spectrum of the Brillouin mode showing an intrinsic Q factor of 63.3 million. (c) Typical transmission spectrum of the microresonator. Brillouin mode and Pump mode are indicated. (d) The transmission power spectra of the pump laser, comb, and the backward laser (including reflected pump laser, Brillouin laser, and comb) at a pump power of 141 mW. The Brillouin laser first turns on which subsequently acts as a secondary laser to excite the Kerr frequency comb as the pump laser scans. (e) Zoomed-in transmission power spectra of the comb and backward laser, respectively. Typical steps in the transmission power spectrum of the comb indicate that the frequency range of soliton existence is over 200 MHz. Frequency is calibrated by a fiber interferometer. (f) The optical spectrum of backward laser before the comb generation. (g) The radiofrequency (RF) spectrum of the beat note between the reflected pump laser and Brillouin laser with a 20 kHz resolution bandwidth (RBW) at 10.78 GHz center frequency (CF).

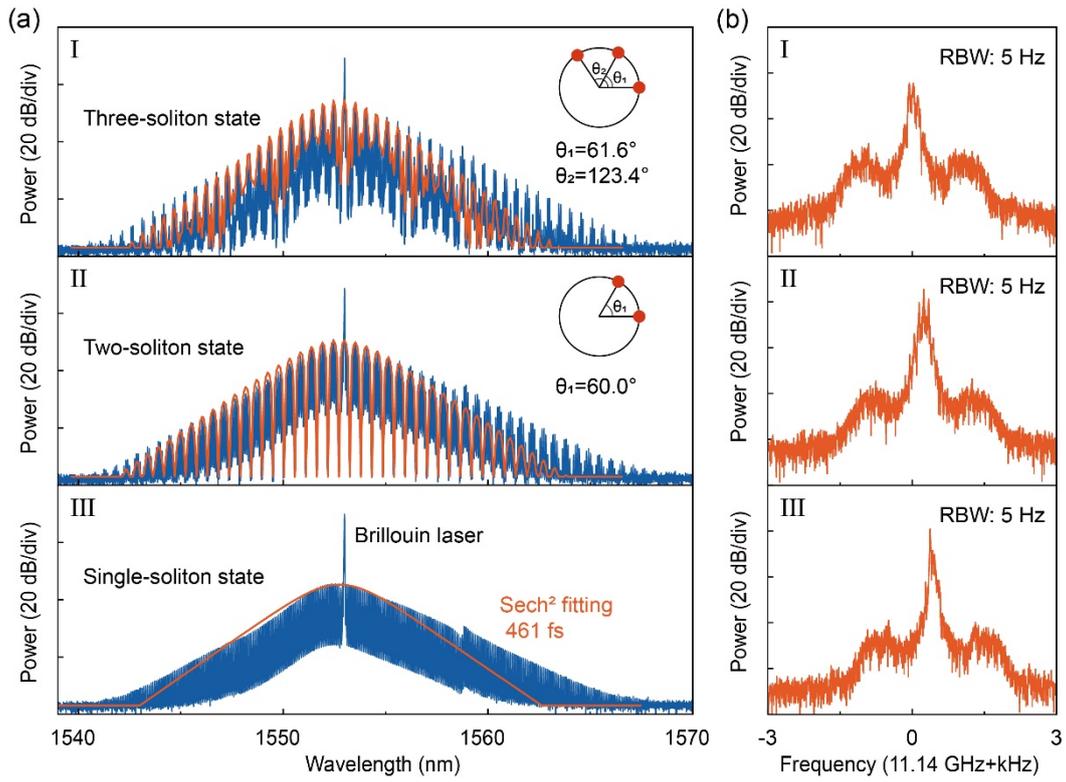

FIG. 3. Brillouin-Kerr soliton frequency comb. (a) The optical spectra of different soliton states. I, II, and III show the measured optical spectra and fitted envelopes of the three-soliton state, two-soliton state, and single-soliton state, respectively. Insets show the relative positions of solitons circulating in the microdisk. (b) The RF spectra of the corresponding soliton repetition frequency for the cases of I, II, and III in (a), measured with 11.14 GHz center frequency and 5 Hz RBW.

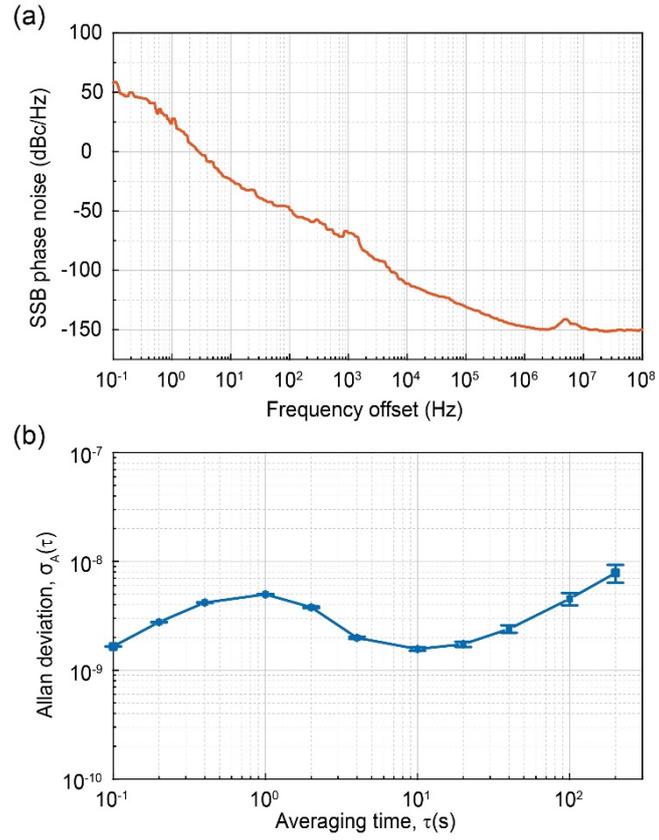

FIG. 4. Phase noise and Allan deviation of the generated RF signal by the Brillouin-Kerr soliton frequency comb. (a) Single sideband (SSB) phase noise of the soliton repetition rate. (b) Allan deviation of the soliton repetition rate measured with 0.1 s gate time.

# Supplementary Material for

# Brillouin-Kerr soliton frequency combs in an optical microresonator

Yan Bai[1]†, Menghua Zhang[1]†, Qi Shi[1]†, Shulin Ding[1]†, Zhenda Xie[1], Xiaoshun Jiang[1]*, and Min Xiao[1,2]

[1]National Laboratory of Solid State Microstructures, College of Engineering and Applied Sciences, and School of Physics, Nanjing University, Nanjing 210093, China.
[2]Department of Physics, University of Arkansas, Fayetteville, Arkansas 72701, USA.
†These authors contributed equally.
*Email: jxs@nju.edu.cn.

## I. Theoretical Model

To model the Brillouin-Kerr frequency comb in an optical microresonator, we employ the coupled mode equations by combining the nonlinear four-wave-mixing (FWM) and the simulated Brillouin scattering (SBS) processes as well as the interaction between them. Under the rotating-wave approximation, for our system, after some algebra we have obtained the following set of coupled-mode equations [1,2]:

$$\frac{da_+}{dt} = \left(-i\Delta\omega_p - \frac{\gamma_+}{2}\right)a_+ - ig_b a_{-,0} b + \sqrt{\kappa_+} s_{in}, \tag{S1}$$

$$\frac{da_{-,n}}{dt} = \left(-i\Delta\omega_{b,n} - \frac{\gamma_-}{2}\right)a_{-,n} - i\delta_0 g_b a_+ b^* - ig_2 \sum_{k,l,m} \delta_{n-(k-l+m)} a_{-,k} a_{-,l}^* a_{-,m}, \tag{S2}$$

$$\frac{db}{dt} = \left(-i\Delta\Omega_m - \frac{\gamma_m}{2}\right)b - ig_b a_+ a_{-,0}^*. \tag{S3}$$

Here, $t$ is the slow time compared to the cavity roundtrip, $a_+$ and $b$ respectively stand for amplitudes of the pump optical mode and the acoustic mode, $a_{-,0}$ and $a_{-,n}$ are, respectively, the amplitudes of the Brillouin and the comb modes with index $n$, $\Delta\omega_p = \omega_p - \omega_+$ and $\Delta\Omega_m = \Omega_m - \Omega_0$ are, respectively, the pump frequency detuning and the frequency shift of the acoustic modes, $\Omega_m$, $\omega_+$, and $\Omega_0$ respectively represent the acoustic wave frequency, the pump and acoustic mode resonant frequencies, $\Delta\omega_{b,0} = \omega_{b,0} - \omega_{-,0}$ and $\Delta\omega_{b,n} = \omega_{b,n} - \omega_{-,n}$ are, respectively, the frequency shifts of the Brillouin mode and the comb modes with $\omega_{b,0}$, $\omega_{b,n}$, $\omega_{-,0}$, and $\omega_{-,n}$ being the Brillouin and comb wave frequencies and Brillouin and comb mode resonant frequencies, $\gamma_m$ is the decay rate of the acoustic mode, $\gamma_+ = \gamma_{+,0} + \kappa_+$ and $\gamma_- = \gamma_{-,0} + \kappa_-$ denote the total decay rates of the pump mode and the Brillouin mode in terms of the intrinsic decay rate $\gamma_{\pm,0}$ and the external coupling rates to the waveguide $\kappa_\pm$, respectively, $g_b$ is the coupling coefficient between the optical and acoustic modes in the SBS process, $g_2$ is the Kerr nonlinear coupling coefficient in the FWM process, $\delta_n$ is the Kronecker delta and $|s_{in}|^2$ represents the input pump power. To describe the SBS process, we use the model of three-mode coupled equations [1], which describes the interaction of two optical waves interaction with an acoustic wave. We also build a model with the modal-expansion formalism in an optical microresonator to describe the generation of the Kerr comb with the intracavity Brillouin laser [2], where the interaction modes are labeled by $n$, $k$, $l$, and $m$ with the relation $\omega_n + \omega_l \leftrightarrow \omega_k + \omega_m$. Note that, in the model, we assume all comb modes have the same intrinsic decay rates and external coupling rates for the sake of simplicity. The resonant frequency of the comb mode with index $n$ can be described through the relation $\omega_{-,n} = \omega_{-,0} + \sum_{j \geq 1} \frac{1}{j!} D_j n^j$ (Here, $D_1$ is the free spectral range and $D_k$ is the $k$-th order dispersion in Taylor series for eigenfrequencies around the Brillouin mode $\omega_{-,0}$ of the cavity). The nonlinear coupling coefficients for the SBS and FWM processes are, respectively, estimated as [3,4]:

$$g_b^2 = \frac{G_b c^2 \gamma_m}{2 n_0^2 V_{eff}}, \tag{S4}$$

$$g_2 = \frac{n_2 c \omega_{0,-}}{n_0^2 V_{eff}}, \tag{S5}$$

where $G_b$ is the bulk Brillouin center-line gain coefficient, $n_0$ and $n_2$ are, respectively, the linear and nonlinear refractive indices of silica, $V_{eff} = A_{eff} L_{eff}$ is the effective mode volume, $L_{eff}$ and $A_{eff}$ are, respectively, used to represent the effective mode length and area, and $c$ is the light speed in vacuum. The output pump and comb waves can be obtained through the input-output relation, which are

$$s_{out,+} = s_{in} - \sqrt{\kappa_+} a_+, \tag{S6}$$

$$s_{out,-} = \sqrt{\kappa_-} \sum_n a_{-,n} e^{i(n D_1 + \Delta \omega_{b,n})t}, \tag{S7}$$

respectively.

## II. Brillouin Laser with Kerr self-phase modulation

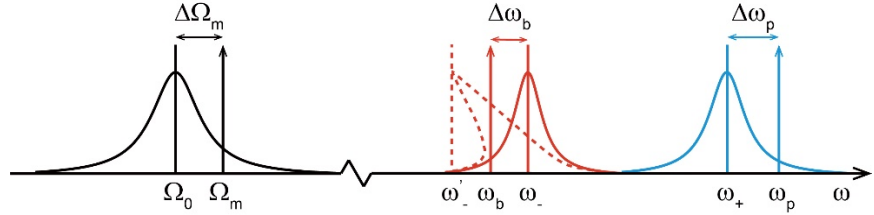

FIG. S1. Illustration of Brillouin laser with Kerr self-phase modulation. $\omega_-$ stands for $\omega_{-,0}$, $\omega'_- = \omega_{-,0} + \Delta\omega'_{b,0}$, and $\Delta\omega_b$ stands for $\Delta\omega_{b,0} = \Delta\omega'_{b,0} + \Delta\omega''_{b,0}$. Dashed red curve, cavity response of Brillouin mode after Kerr self-phase modulation.

To study the property of the Brillouin laser before the generation of the optical parametric oscillation (OPO), we set $a_{-,n \neq 0} = 0$ in Eq. (S2) to have:

$$\frac{da_{-,0}}{dt} = \left( -i \Delta \omega_{b,0} - \frac{\gamma_-}{2} \right) a_{-,0} - i g_b a_+ b^* - i g_2 |a_{-,0}|^2 a_{-,0}. \tag{S8}$$

Then, Eqs. (S1), (S3), and (S8) can be used to describe the Brillouin lasing after considering the self-phase modulation induced by the Kerr nonlinearity. Here, we focus on two kinds of analytical steady state solutions. By setting the ansatz $a_{-,n} = 0$ in Eqs. (S1)-(S3), we can obtain a steady state solution before the Brillouin lasing occurring:

$$a_+ = \sqrt{\kappa_+} s_{in} / \left( i \Delta \omega_p + \frac{\gamma_+}{2} \right). \tag{S9}$$

Another kind of solutions are the equilibriums with Brillouin laser. By further setting $\frac{da_+}{dt} = \frac{da_{-,0}}{dt} = \frac{db}{dt} = 0$ and noting the phase matching condition $\Omega_m = \omega_p - \omega_b$ in SBS, we find the Brillouin laser will satisfy the following relations as an equilibrium between loss and gain with balance of the frequency shift terms in Eq. (S8) (note that the exact solutions can be further solved by treating $\Delta\Omega_m$ as variable and eliminating $\Delta\omega_{b,0}, a_{-,0},$ and $a_+$):

$$\gamma_- = \frac{\gamma_m G}{\Delta\Omega_m^2 + \gamma_m^2}|a_+|^2, \tag{S10}$$

$$\Delta\omega_{b,0} = \Delta\omega'_{b,0} + \Delta\omega''_{b,0}, \tag{S11}$$

Where $G = g^2$, $\Delta\omega'_{b,0} = -\frac{\gamma_m}{\gamma_m+\gamma_-}g_2|a_{-,0}|^2$, $\Delta\omega''_{b,0} = \frac{\gamma_-}{\gamma_m+\gamma_-}(\Delta\omega_0 + \Delta\omega_p)$, and $\Delta\omega_0 = \omega_+ - \omega_{-,0} - \Omega_0$ is the resonant frequency mismatch among the three modes. After substituting Eq. (S9) into Eq. (S10), the threshold of the Brillouin laser can be evaluated subject to the following condition:

$$\frac{16\gamma_m G \kappa_+ |s_{in}|^2}{(4\Delta\Omega_m^2 + \gamma_m^2)(4\Delta\omega_p^2 + \gamma_+^2)} = \gamma_-. \tag{S12}$$

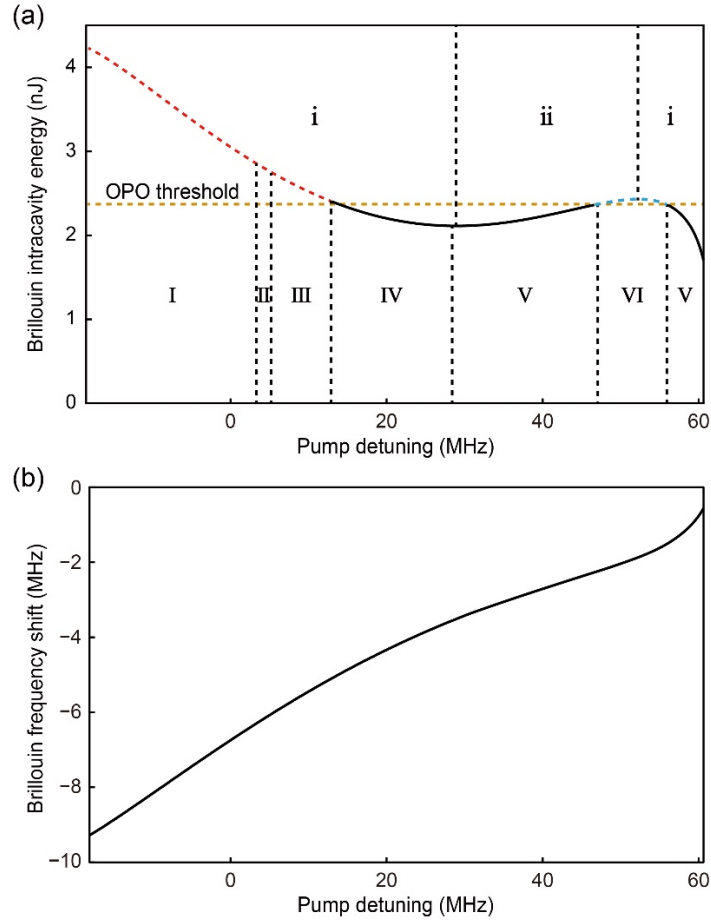

FIG. S2. The intracavity energy (a) and frequency shift (b) of the Brillouin wave as a function of the input pump detuning. Black (red and blue) solid (dashed) curves in (a) are steady states below (beyond) OPO threshold. Capital Roman numerals in (a) label the pump detuning regions with different evolution results of Brillouin-Kerr frequency comb: (I) chaotic combs; (II) soliton breathers; (III) perturbation generated solitons; (IV) solitons and flat states; (V) flat states; (VI) Turing patterns. The thermal stability of the generated Brillouin lasing versus the frequency detuning of the input pump is labeled with lower-case Roman numerals in (a). The Brillouin laser in region (i) is thermally stable while in region (ii) is thermally unstable. The cavity parameters are $V_{eff} = 7.5 \times 10^{-13}$ m$^3$ with $A_{eff} = 40$ μm$^2$ and $L_{eff} = 2\pi \times 3$ mm, $\gamma_{+,0} = 2\pi \times 4$ MHz, $\gamma_{-,0} = 2\pi \times 3$ MHz, $\kappa_+ = 2\pi \times 30$ MHz, $\kappa_- = 2\pi \times 1.5$ MHz, $\gamma_m = 2\pi \times 50$ MHz, and $\omega_0 = -2\pi \times 48$ MHz. The input pump power is $|s_{in}|^2 = 150$ mW.

As shown in Fig. S2, we give an example of the simulated intracavity energy and frequency shift of the Brillouin wave as a function of the input pump detuning according to the solutions of Eqs. (S10) and (S11). Here, we set the bulk Brillouin center-line gain coefficient $G_b = 3.46 \times 10^{-12}$ m/W, the refractive index $n_0 = 1.45$, and the nonlinear refractive index $n_2 = 2.6 \times 10^{-20}$ m$^2$/W for a silica microresonator. Note that we apply $G_b = \frac{\gamma_e^2 \omega_{-,0}^2}{2\pi n_0 v c^3 \rho \gamma_m}$ [5,6] to estimate this parameter, where $\gamma_e = 1.5$ is the electrostrictive constant [1] , $v = 5664$ m/s is the speed of sound, and $\rho = 2200$ kg/m$^3$ is the density. The other parameters used in the calculations are listed in the caption of Fig. S2. As shown in Fig. S2(b), one can find that for a certain input pump power, the generated Brillouin laser can be red detuned with both blue- and red-detuned input pump. We believe this finding is of interest for Kerr soliton generation.

### III. OPO generation with intracavity Brillouin laser

To investigate the OPO with the intracavity Brillouin laser, for simplicity, we assume the dispersion relation $\omega_{-,n} \approx \omega_{-,0} + D_1 n + \frac{D_2}{2} n^2$ by keeping only the linear and second order terms but neglecting all other higher order terms. To find the threshold of OPO, a well-known method is to study the linear stability of the mode pair among the comb modes at their trivial equilibrium $a_{-,\pm l} = 0$ [2]. The specific perturbation pairs $\delta a_{-,\pm l}$, respectively, are then added into the equilibrium. By assuming the Brillouin mode $a_{-,0}$ with a frequency shift $\Delta \omega_{b,0}$ being still at its equilibrium, we eventually find two dynamical equations for the mode perturbation pair:

$$\frac{d\delta a_{-,l}}{dt} = \left(-i\Delta\omega_{b,0} - \frac{D_2}{2} l^2 - \frac{\gamma_-}{2}\right) \delta a_{-,l} - 2ig_2 |a_{-,0}|^2 \delta a_{-,l} - ig_2 a_{-,0}^2 \delta a_{-,-l}^*, \quad (S13)$$

$$\frac{d\delta a_{-,-l}}{dt} = \left(-i\Delta\omega_{b,0} - \frac{D_2}{2} l^2 - \frac{\gamma_-}{2}\right) \delta a_{-,-l} - 2ig_2 |a_{-,0}|^2 \delta a_{-,-l} - ig_2 a_{-,0}^2 \delta a_{-,l}^*, \quad (S14)$$

where $\delta a_{-,\pm l}^*$ are the conjugate pair of $\delta a_{-,\pm l}$. By further considering the dynamical equations of $\delta a_{-,\pm l}^*$, we find an eigen problem with a fourth-order matrix:

$$\frac{d\delta A_l}{dt} = M \delta A_l, \quad (S15)$$

where, $M = -\frac{\gamma_-}{2} I - i \begin{bmatrix} B & C \\ -C^* & -B^* \end{bmatrix}$ with $I$ being the identity matrix, $\delta A_l = [\delta a_{-,l}, \delta a_{-,-l}, \delta a_{-,l}^*, \delta a_{-,-l}^*]^T$, $B = \begin{bmatrix} \Delta\omega_{b,0} + \frac{D_2}{2} l^2 + 2g_2 E_{-,e} & 0 \\ 0 & \Delta\omega_{b,0} + \frac{D_2}{2} l^2 + 2g_2 E_{-,e} \end{bmatrix}$, and $C = \begin{bmatrix} 0 & g_2 a_{-,0}^2 \\ g_2 a_{-,0}^2 & 0 \end{bmatrix}$. $E_{-,e} = |a_{-,0}|^2$ is the intracavity energy of the Brillouin mode at the equilibrium. The two degenerated eigenvalues of Eq. (S15) are:

$$-\frac{\gamma_-}{2} \pm \sqrt{g_2^2 E_{-,e}^2 - (\Delta\omega_{b,0} + \frac{D_2}{2} l^2 + 2g_2 E_{-,e})^2}. \quad (S16)$$

Eq. (S16) can be used to investigate the generation of the primary comb or the number of rolls in Turing patterns as done in the previous works of Kerr comb generation [2,7]. When the real parts of all (one of) eigenvalues are negative (positive), the perturbation pair $\delta a_{-,\pm l}$ will decay (diverge) to zero (infinity) as time passes. The OPO threshold of the perturbation pair is defined as the separating point of these two opposite cases with $\frac{\gamma_-}{2} = \sqrt{g_2^2 E_{-,e}^2 - (\Delta\omega_{b,0} + \frac{D_2}{2} l^2 + 2g_2 E_{-,e})^2}$.

As a result, to generate the Kerr combs, the system needs to be operated above the threshold and to satisfy the condition $E_{-,e} \geq \frac{\gamma_-}{2g_2}$. Although more than one perturbation pair $\delta a_{-,\pm l}$ may be beyond the threshold, the one with the largest gain is likely to excite from noise. Subsequently, we intriguingly find two cases in which the perturbation pair $\delta a_{-,\pm l}$ possess the largest gain. In one case, since the resonator satisfies the anomalous dispersion ($D_2 < 0$), when $\Delta\omega_{b,0} \geq -2g_2 E_{-,e} - \frac{D_2}{2}$ [see the blue dashed curve in Fig. S2(a)], the perturbation pair with $l$ close to $\sqrt{\frac{2(\Delta\omega_{b,0}+2g_2 E_{-,e})}{D_2}}$ are likely to be excited [2]. The primary comb with mode number $l > 1$ will be generated from perturbation and the Turing patterns will be observed [7]. In the other case, when $\Delta\omega_{b,0} < -2g_2 E_{-,e} - \frac{D_2}{2}$ and for $l \geq 1$, the number of the perturbation pair $\delta a_{-,\pm l}$ with the largest gain will always be $l = 1$. The frequency shift of the Brillouin wave in the case satisfies $-\sqrt{g_2^2 E_{-,e}^2 - \frac{\gamma_-^2}{4}} - \frac{D_2}{2} - 2g_2 E_{-,e} \leq \Delta\omega_{b,0} \leq \sqrt{g_2^2 E_{-,e}^2 - \frac{\gamma_-^2}{4}} - \frac{D_2}{2} - 2g_2 E_{-,e}$. In this case [see the red dashed curve in Fig. S2(a)], we numerically (see simulation details in the section V) find the primary combs with mode number $l = 1$ will be generated from perturbation, while the chaotic combs, soliton breathers, or solitons [see region I, II, and III in Fig. S2(a)] could be directly excited in the related regions.

### IV. Thermal stability of intracavity Brillouin laser

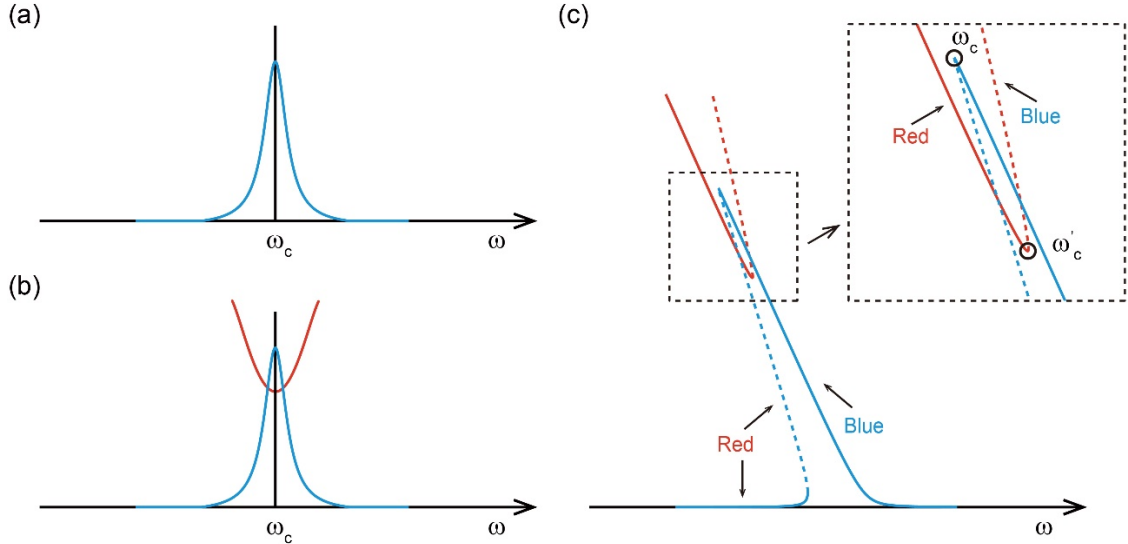

FIG. S3. Illustration of total intracavity energy transition with TOE. For simplicity, here we set $\Delta\omega_0 = 0$. $\omega_c$ is the resonance frequency of the pump mode. (a) Steady states of a cavity mode with Lorentz lineshape. (b) State transition in a Brillouin laser under a large power pump, curves: red, steady states of Brillouin laser; blue, states before Brillouin lasing. (c) Brillouin laser with TOE, curves: red, steady states of Brillouin laser; blue, states before Brillouin lasing; solid, stable states; dashed, unstable states. $\omega_c'$, resonant pump point of Brillouin laser. The words "red" and "blue" indicate red and blue detuned pump states.

To study the thermal stability of the Brillouin laser, we further consider the thermal-optic effect (TOE) due to the material absorption [8]. Then, the thermal dynamics of the generated Brillouin lasing can be modeled by replacing $\Delta\omega_p$ and $\Delta\omega_{b,n}$ with $\Delta\omega_p - \beta_+\Delta T$ and $\Delta\omega_{b,n} - \beta_{-,n}\Delta T$ in Eqs. (S1) and (S2), respectively, and adding another equation to describe the dynamics of cavity temperature difference $\Delta T$ with the surroundings [9]:

$$\frac{d\Delta T}{dt} = -\frac{1}{\tau_T}\Delta T + \alpha_+|a_+|^2 + \alpha_{-,n}\sum_n|a_{-,n}|^2. \quad (S17)$$

Here, $\beta_+$ and $\beta_{-,n}$ are, respectively, the thermal induced frequency shifts of the pump and comb modes, $\tau_T$ is the thermal relaxation time of the cavity, and $\alpha_+$ and $\alpha_{-,n}$ respectively represent the absorption coefficients related to the TOE of the pump and comb modes. To analysis the stability property of the steady states of the Brillouin laser, we can neglect the FWM process to ease the problem. As such, Eq. (S17) reduces to:

$$\frac{d\Delta T}{dt} = -\frac{1}{\tau_T}\Delta T + \alpha_+|a_+|^2 + \alpha_{-,0}|a_{-,0}|^2. \quad (S18)$$

In a typical cavity without the thermal effect, the total intracavity energy of steady states versus the pump detuning is shown in Fig. S3(a), which exhibits a Lorentzian lineshape. At this condition, the system is stable with both blue- and red-detuned pump. However, for high power input pump in high-Q microcavity, only the blue detuned pump is stable while the red-detuned pump is unstable [see blue dashed curve in Fig. S3(c)] due to the TOE induced bistability [8]. Yet, these states will alter [see red curve in Fig. S3(b) in case of $\Delta\omega_0 = 0$] due to the generation of the Brillouin laser [10]. Finally, the steady states considering TOE are shown in Fig. S3(c). From the red curves in Fig. S3(c), we find the Brillouin laser are stable at certain red-detuned input pump. This leads to the conclusion that the thermally stable region will be extendable to red pump detuning frequency with a high power pump Brillouin laser scheme in a silica microcavity.

As an example with $\Delta\omega_0 < 0$ [see Fig. S2(a)], we further divide the calculated steady states of the Brillouin laser into thermally stable (region i) and unstable (ii) regions. From Fig. S2(a), we learn that the Brillouin laser is thermally stable for the blue and part of the red detuned input pump, which is slightly different from the case of $\Delta\omega_0 = 0$. From the above study, we find that the generated Brillouin laser with red detuning is thermally stable, which gives an excellent platform for generating Kerr solitons.

V. Brillouin-Kerr soliton formation

In Fig. S2(a), we have calculated and labeled the evolution results of different steady states of the Brillouin laser by further including the FWM process. These simulations rely on the numerical solutions of Eqs. (S1)-(S3). The numerical calculations are performed with the fourth-order Runge-Kutta method, where 512 modes are used as the comb modes. Here, the second order dispersion parameter $D_2 = 2\pi \times 10$ kHz and all higher order dispersions are neglected. The random vacuum fluctuation fields serve as the initial seeds for the SBS and FWM processes. Moreover, we also employ a computation-efficient algorithm to calculate the cubic nonlinear terms, $g_2\sum_{k,l,m}\delta_{n-(k-l+m)}a_{-,k}a^*_{-,l}a_{-,m}$ [11]. As shown in Fig. S4, we have theoretically demonstrated the symmetric three solitons generation from the noise by setting the pump detuning $\Delta\omega_p = 2\pi \times 12$ MHz. To further achieve the single soliton state, we can employ the backward tuning technology to reduce the number of the solitons [12], which has been demonstrated in the experiment.

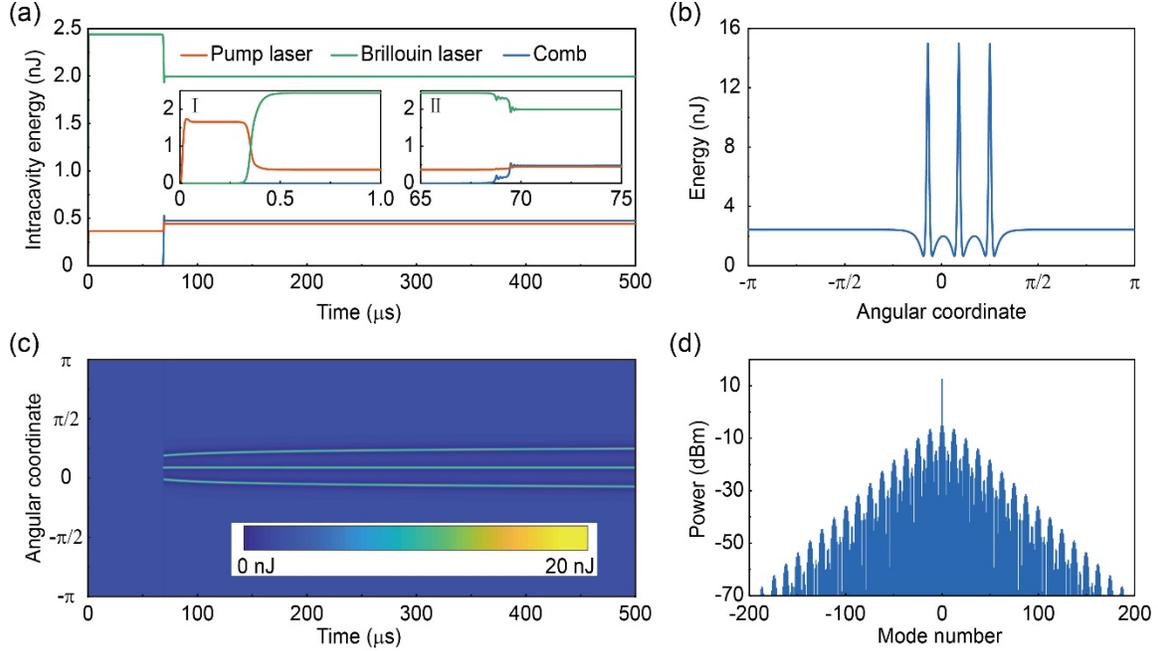

FIG. S4. Three solitons formation. (a) Intracavity energy evolution, insets: I and II are zoom-in display of the Brillouin laser and the three solitons generation, respectively. (b) Final energy distribution of comb modes in angular coordinate. (c) Energy distribution evolution of comb modes in angular coordinate. (d) Simulated optical spectrum of the three-soliton comb.

By adding the thermal dynamics described by Eq. (S17) into the calculation of Eqs. (S1)-(S3), we can also theoretically show that the generated Brillouin-Kerr solitons are thermally stable. Here, the stability of Brillouin-Kerr solitons can be understood as following: In the presence of the Brillouin-Kerr solitons, the cavity resonance will experience a red shift due to heat accumulation in soliton energy growth. This red-shifted cavity resonance in turn results in a blue shift of the Brillouin laser, which adversely constrains the growth of the soliton energy. When the growth and the suppression of the soliton energy are balanced to each other, a thermally stable soliton state will be finally obtained.

## VI. Experimental setup

The experimental setup for the intracavity Brillouin laser pumped soliton generation is shown in Fig. 1(c). The input pump laser at wavelength 1550 nm (emitted by a tunable external cavity diode laser with a short-term linewidth of ~ 10 kHz) is first amplified by an erbium doped fiber amplifier (EDFA), and then coupled into the silica microdisk through a fiber taper. A variable optical attenuator (VOA) and a fiber optic polarizer (FPC) are subsequently used to adjust the pump power and optimize the polarization state of the laser, respectively. The tunable bandpass filter (TBF) is used to filter out the noise from the EDFA. The power meter (PM) is used to monitor the input pump power. The transmission spectrum of the pump light is measured by a slow photodiode (PD1) and monitored by the oscilloscope (OSC). The backward signals, including the generated Brillouin laser, the Kerr frequency comb and the reflected pump light due to the backscattering, are extracted by an optical fiber circulator (C1) and measured by an optical spectrum analyzer (OSA) with 0.02 nm resolution. To measure the microwave signal generated by the Brillouin-Kerr frequency comb, the backward laser lights are routed by circulator 1 (C1), Coupler 2, Coupler 3, circulator 2 (C2), fiber Bragg grating (FBG), Coupler 4 and the optical switch (S), which is then detected through the fast PD3 and analyzed by an electrical spectrum analyzer (ESA). The power of the generated Brillouin-Kerr frequency comb is also measured by a slow detector (PD2). The fiber Bragg gating (FBG) with 1 nm bandwidth is used to filter out the pump laser as well as the generated Brillouin laser. Also, the reflected signal (mainly consisting of the Brillouin laser, the backscattered input pump laser and several comb lines) from the FBG (after passing C2) is further analyzed by a fast photodiode (PD3) and an electrical spectrum analyzer (ESA) to measure the beat notes of the Brillouin laser and the backscattered pump laser as well as the Brillouin-Kerr frequency combs. In the experiment, we can observe the two RF beat signals simultaneously when the Brillouin-Kerr solitons are generated.